# Beyond the *E*-value: stratified statistics for protein domain prediction

Alejandro Ochoa[1,3], ochoa@princeton.edu

John D. Storey[1,3], jstorey@princeton.edu

Manuel Llinás[4], manuel@psu.edu

Mona Singh[2,3]*, mona@cs.princeton.edu

[1]Department of Molecular Biology,

[2]Department of Computer Science, and

[3]Lewis-Sigler Institute for Integrative Genomics, Princeton University, Princeton, New Jersey 08544, USA

[4]Department of Biochemistry and Molecular Biology, and the Huck Institutes of the Life Sciences, Pennsylvania State University, University Park, Pennsylvania 16802, USA

*Corresponding Author

**ABSTRACT**

*E*-values have been the dominant statistic for protein sequence analysis for the past two decades: from identifying statistically significant local sequence alignments to evaluating matches to hidden Markov models describing protein domain families. Here we formally show that for "stratified" multiple hypothesis testing problems—that is, those in which statistical tests can be partitioned naturally—controlling the local False Discovery Rate (lFDR) per stratum, or partition, yields the most predictions across the data at any given threshold on the FDR or *E*-value over all strata combined. For the important problem of protein domain prediction, a key step in characterizing protein structure, function and evolution, we show that stratifying statistical tests by domain family yields excellent results. We develop the first FDR-estimating algorithms for domain prediction, and evaluate how well thresholds based on *q*-values, *E*-values and lFDRs perform in domain prediction using five complementary approaches for estimating empirical FDRs in this context. We show that stratified *q*-value thresholds substantially outperform *E*-values. Contradicting our theoretical results, *q*-values also outperform lFDRs; however, our tests reveal a small but coherent subset of domain families, biased towards models for specific repetitive patterns, for which FDRs are greatly underestimated due to weaknesses in random sequence models. Usage of lFDR thresholds outperform *q*-values for the remaining families, which have as-expected noise, suggesting that further improvements in domain predictions can be achieved with improved modeling of random sequences. Overall, our theoretical and empirical findings suggest that the use of stratified *q*-values and lFDRs could result in improvements in a host of structured multiple hypothesis testing problems arising in bioinformatics, including genome-wide association studies, orthology prediction, motif scanning, and multi-microarray analyses.

**INTRODUCTION**

The evaluation of statistical significance is crucial in nearly all genome-wide studies, including detecting differentially expressed genes in microarray or proteomic studies, performing genome-wide association studies, and uncovering significant matches in homology searches. Different biological applications have settled for different statistics to set thresholds on. In the field of biological sequence analysis, accurate statistics for pairwise alignments and their use in database search [1–3] were introduced with the use of random sequence models and *E*-values two decades ago [4,5]. Sequence similarity searches have evolved further, from the pairwise comparison tools of FASTA [3] and BLAST [5], to sequence-profile [6–8] and profile-profile [9–12] comparisons. While the different approaches to detect sequence similarity have relied on a variety of statistics, including bit scores [13,14] and *Z*-scores [3], most modern approaches are based on *E*-values.

Detecting sequence similarity in order to uncover homologous relationships between proteins remains the single most powerful tool for function prediction. A key concept in many modern techniques for detecting sequence similarity is that of domains, the fundamental units of protein structure, function, and evolution. Homologous domains are grouped into "families" that may be associated with specific functions and structures, and these domain families define an organization of protein space. Each domain family is typically modeled with a profile or a hidden Markov model (HMM) [13,15], and there are many databases of domain HMMs, each providing a different focus and organization of domain space, including Pfam [14], Superfamily [16], and Smart [17]. Although HMM-based software, such as the state-of-the-art HMMER program [18], has features that make it superior in remote homology detection to its predecessors, accurate statistics for evaluating significance arose only recently [19].

At its core, the protein domain prediction problem is a multiple hypothesis testing problem, where tens of thousands of homology models (one for each domain) are scored against tens of millions of sequences. Each comparison yields a score $s$ and a $p$-value, defined as the probability of obtaining a score equal to or larger than $s$ if the null hypothesis holds. While a small $p$-value threshold (for example, 0.05 or smaller) is acceptable to declare a single statistical test significant, this is inappropriate when a large number of tests are performed. Instead, as in other sequence analysis applications, thresholds for domain predictions are typically based on the $E$-value. The $E$-value can be computed from a $p$-value threshold $p$ as $E=pN$, where $N$ is the number of tests, and yields the expected number of false positives at this $p$-value. While setting $E$-value thresholds makes sense in the context of a single database search, especially when few positives are expected, they are less meaningful when millions of positives are obtained, and hence a larger number of false positives might be tolerated. Moreover, in multiple database search problems, such as BLAST-based orthology prediction [20] or genome-wide domain prediction [21], $E$-values are usually not valid because many searches are performed without the additional multiple hypothesis correction required.

Though the $E$-value has thus far been the default statistic used to detect sequence similarity, recent work has demonstrated that False Discovery Rate (FDR) control is a powerful approach for multiple hypothesis testing [22]. The FDR is loosely defined as the proportion of all significant tests that are expected to be false, and in practice is estimated as the familiar $E$-value normalized by the number of predictions made. The FDR does not increase with database size $N$ the way the $E$-value does; thus, predictions do not usually lose significance with the FDR as the database grows. The FDR also does not require additional correction in the case of multiple database queries. The FDR can be computed from $p$-values using the Benjamini-Hochberg procedure [22]. The $q$-value statistic is the

FDR-analog of the *p*-value, and it provides conservative and powerful FDR control [23]. The *q*-value of a statistic *t* is the minimum FDR incurred by declaring *t* significant [23]. Thus, *q*-values vary monotonically with *p*-values, and they are easily estimated from *p*-values [23]. While *E*-values control the number of false positives, *q*-values control their proportion. The local FDR (lFDR) measures the proportion of false positives in the infinitesimally small section of the data that lies around the chosen threshold, and hence it is a "local" version of the FDR [24]; it is also equivalent to the Bayesian posterior probability that a prediction is false [24]. However, *q*-value estimates are much more robust than lFDR estimates, since the former are based on empirical cumulative densities, which converge uniformly to the true cumulative densities [25,26]. On the other hand, lFDR estimates are local fits to the density, so they are comparably much more susceptible to noise, especially on the most significant tail of the distribution. The FDR [22], *q*-value [23], and lFDR [24] have all been successfully used in many areas of bioinformatics, including gene expression microarray analysis [24,27,28], genome-wide association studies (GWAS) [27,29], and proteomics analysis [30–34].

Here we introduce the first FDR- and lFDR-estimating algorithms for domain prediction. An essential feature of our approach is that our statistical tests are stratified, rather than pooled, across domains families. We prove that stratified problems, such as domain prediction, are optimally tackled using the lFDR statistic. For domain prediction, we evaluate how well thresholds based on stratified lFDRs and *q*-values perform using five independent approaches for estimating empirical FDRs and empirical *E*-values in this context. Through extensive benchmarking using the Pfam database and the HMMER domain prediction software, we show that the use of stratified *q*-values increases domain prediction by 6.7% compared to the Standard Pfam on UniRef50 [35]. In contrast to theory, we also find that in practice *q*-values outperform lFDRs. Further, we find that while the empirical FDRs for most domain families are similar to what is expected based on our *q*-value thresholds, some families tend to have larger FDRs, suggesting that the standard null model is inappropriate for them and that the initial *p*-values estimated for their matches are inaccurate. Specifically, domain families with larger-than-expected empirical FDRs are enriched for those containing repetitive patterns, such as coiled-coils, transmembrane domains, and other low-complexity regions. When only families with as-expected FDRs are considered, the use of *q*-values increases domain prediction by 8.8% compared to the Standard Pfam, and lFDRs further outperform *q*-values, suggesting that further performance improvements are possible for domain identification if statistical modeling of these families is improved in the future.

We note that stratified FDR analyses have been previously explored [36–39], and have been successfully applied to GWAS in particular [29,40,41]. Thus, the same statistical solution we introduce for domain recognition applies to a wide variety of problems in which many statistical tests can be analyzed separately, including GWAS (stratifying by candidate or genic regions), orthology prediction (stratifying by each ortholog database search), motif scanning (stratifying by each motif search across a genome), multi-microarray analysis (stratifying by each microarray), and other multi-dataset analyses. Overall, we expect our theoretical results, advocating for the use of stratified $q$-values and lFDRs, to have impact on many other applications in bioinformatics and beyond.

**RESULTS**

**FDR definitions.** We briefly review the relevant FDR definitions; for a comprehensive overview, see [42]. The FDR definitions are in terms of two random variables: $V$ is the number of false positive predictions, and $R$ is the total number of significant tests [22]. These quantities are usually parametrized in terms of the $p$-value threshold $t$, which in the case of independent $p$-values gives us expected values of

$$\mathrm{E}[\,V(t)\,] = t\,\pi_0\,N,$$
$$\mathrm{E}[\,R(t)\,] = F(t)\,N,$$

where $\pi_0$ is the proportion of tests which are truly null, $N$ is the total number of tests, and $F(t)$ is the cumulative density of $p$-values [23,24,43]. Note that $\mathrm{E}[\,V(t)\,]$ gives the $E$-value statistic.

There are two closely-related versions of the FDR that we use in our work: the positive FDR (pFDR) and marginal FDR (mFDR) [42,43]. These are defined as:

$$\mathrm{pFDR} = \mathrm{E}\left[\left.\frac{V}{R}\right|R>0\right],$$
$$\mathrm{mFDR} = \frac{\mathrm{E}[V]}{\mathrm{E}[R]}.$$

The advantages of the pFDR compared to the original FDR=E[V/R] definition of Benjamini and Hochberg [22] are discussed in [43]. Further, if $p$-values are drawn independently from a two-component distribution (of null and alternative hypotheses; **Figure 1**), it has been shown that the pFDR and mFDR are equivalent and can be expressed as the "Bayesian FDR" [24,44], which is a posterior probability:

$$\mathrm{pFDR} = \mathrm{mFDR} = \Pr(H=0 | p \leq t) = \frac{t\,\pi_0}{F(t)},$$

where $H=0$ denotes that the null hypothesis holds [43]. The pFDR and mFDR are also asymptotically equal under certain forms of "weak dependence," as defined in [45]. We note that our domain prediction problem satisfies both very large sample sizes and very weak dependence: our dataset contains millions of protein sequences and thousands of HMMs, and we only expect dependent null $p$-values from very similar sequences and the same or similar HMMs. These are very small subsets of all hypotheses tested, even on each stratum (that is, for any one HMM). For this reason, we use the term FDR to refer loosely to all of the FDR definitions.

The local FDR (lFDR) is the Bayesian posterior error probability (PEP) derived from the Bayesian FDR [24], and it is related to our FDR parameters by

$$\text{lFDR}(t) = \frac{\pi_0}{f(t)},$$

where $f(t) = F'(t)$ is the $p$-value density at $t$. Thus, while the Bayesian FDR is a ratio of areas, the lFDR is a ratio of densities (**Figure 1**) [44].

The $q$-value of a statistic $t$ (usually, but not necessarily, a $p$-value) is the minimum pFDR incurred by declaring $t$ significant [23]. Estimated $q$-values are efficiently constructed from $p$-values, and these estimates control the pFDR [23]. Estimated $q$-values are based on the Bayesian FDR formula, by estimating $\pi_0$ and $F(t)$; our lFDR estimates follow similarly but $f(t)$ has to be estimated instead. See **Supplementary Methods** for the algorithms for estimating $q$-values and lFDRs.

**Equal stratified lFDR thresholds maximize predictions while controlling the combined FDR.** We first prove a theorem regarding how to optimally choose thresholds for stratified problems in general. (Recall that in the specific case of domain prediction, each domain family defines a stratum.) We wish to find $p$-value thresholds $t_i$ per stratum $i$ that maximize the number of predictions across strata constrained to some maximum FDR of the combined strata. We will show that in the optimal solution the stratified lFDRs must be equal. This result is consistent with the related Bayesian classification problem, where the lFDR is known to be optimal [43].

Let us define the Bayesian FDR model quantities $N_i$, $\pi_{0,i}$, $F_i(t_i)$ and $f_i(t_i)$ separately for each stratum $i$. We desire to maximize the expected number of predictions across strata

$$\sum_i F_i(t_i) N_i,$$

while constraining the "combined" Bayesian FDR, defined here as the sum of expected false positives across strata divided by the total number of expected predictions, to a maximum value of $Q$, or

$$Q \geq \frac{\sum_i t_i \pi_{0,i} N_i}{\sum_i F_i(t_i) N_i}.$$

This problem can be solved via the equivalent Lagrangian multiplier function $\Lambda$, with the constraint set to strict equality, in a formulation that avoids quotients:

$$\Lambda = \sum_i F_i(t_i) N_i + \lambda \left( \sum_i t_i \pi_{0,i} N_i - Q \sum_i F_i(t_i) N_i \right)$$
$$= \sum_i F_i(t_i) N_i (1 - \lambda Q) + \lambda t_i \pi_{0,i} N_i.$$

Taking the partial derivative of $\Lambda$ with respect to $t_j$, we obtain a necessary condition for optimality,

$$\frac{\partial \Lambda}{\partial t_j} = f_j(t_j) N_j (1 - \lambda Q) + \lambda \pi_{0,j} N_j = 0 \iff$$
$$Q - \frac{1}{\lambda} = \frac{\pi_{0,j}}{f_j(t_j)} = \text{lFDR}_j(t_j),$$

which shows that the lFDR of each stratum must be equal, since the last equation has the same value for every stratum $j$. The same conclusion is obtained if the combined FDR constraint is replaced by a combined $E$-value constraint (**Supplementary Methods**).

**Obtaining $E$-values, $q$-values, and lFDRs for domains.** Each of the 12,273 Pfam domain families was used to scan for domains in each of 3.8 million proteins of UniRef50 (**Supplementary Methods**), for a total of 47 billion statistical tests. The resulting domain predictions are stratified by domain family (HMM), so that each stratum contains $p$-values from which we estimate $q$-values and lFDRs. We note that standard $q$-value and lFDR implementations fail for domain data for two reasons. First, modern HMM software only provides predictions for the smallest $p$-values due to heuristic filters [19]. Second, homologous families (grouped into "superfamilies" [16] or "clans" [14]) produce frequent overlaps that are resolved by removal of all but the most significant match, and thus there are fewer predictions than an independent family analysis would predict, which leads to underestimated FDRs. To address these issues, we remove overlapping domains (keeping those with the smallest $p$-values), and then estimate $q$-values and lFDRs with methods adapted for censored $p$-values (**Methods**). For comparison, we also use $E$-value thresholds and the "Standard Pfam" curated bitscore thresholds (also called "Gathering" or "GA" [14]). Note that a stratified $E$-values approach (separating families) is no different from a combined $E$-value approach in that the ranking of predictions is preserved, since the number of proteins, or tests, is the same per stratum; the stratified $E$-value threshold equals the combined $E$-value threshold divided by the number of strata. Similarly, a combined $q$-value or lFDR approach (obtained by combining the *p*-values of all strata) also preserves the same ordering as a combined *E*-value approach.

**Empirical FDR tests.** We estimate the true FDR via "empirical" FDR tests, to compare all methods on an equal footing, but also to test the accuracy of lFDR and *q*-value estimates in particular. We created or adapted five tests, each of which label domain predictions as either true or false positives (TP, FP) using different statistical and biological criteria. These labels are used to estimate the FDR by computing the proportion of predictions that are FPs.

For simplicity, only two tests are described here in detail and are featured in the main figures. First, the ClanOv ("Clan Overlap") test is based on the expectation that overlapping predicted domains should be related [46]. Pfam annotates related families by placing them into clans, which allows us to label overlapping domains as FPs if they are not in the same clan. In this test, domains are ranked by *p*-value, highest ranking domains are considered as TPs, domains that overlap a higher-ranking domain of the same clan are removed (since they would not be counted as separate predictions), and domains that overlap a higher-ranking domain of a different clan are considered FPs (**Methods, Figure 2B**). Although all FPs in this test would not be predicted by our method when overlaps are removed, this method nevertheless well estimates the amount of noise present in a set of predictions. Second, the ContextC ("Context Coherence") test is based on whether domain pairs predicted within a sequence have been observed together before [47]. In this test, domains are ranked by *p*-value, and the highest ranking domain is always a TP. Subsequently, a domain is a TP if its family has previously been observed with the family of at least one higher-ranking domain, and otherwise it is a FP (**Methods, Figure 2C**). Further, the list of observed family pairs is extended using clans (**Supplementary Methods**).

The principles behind the other three tests are described here briefly: OrthoC ("Ortholog Set Coherence") is based on the expectation that orthologous proteins contain similar domains [48], RevSeq ("Reverse Sequence") estimates noise based on domains predicted on reversed amino acid sequences [49], and MarkovR ("Markov Random") estimates noise based on domains predicted on random sequences generated using a second-order Markov model (**Figure S1**, **Supplementary Methods**).

We can compare methods by asking, at fixed empirical FDR levels, which produce more domain predictions (**Figure 3** and **Figure S2**), more unique families per protein (**Figure S3**), greater amino acid coverage (**Figure S4**), more proteins with predictions (**Figure S5**), and larger "GO information content" scores (as measured using the Gene Ontology [50] and MultiPfam2GO [51], **Supplementary Methods**, **Figure S6**).

**Stratified *q*-values predict more domains than the Standard Pfam, *E*-values, and lFDRs at the same levels of empirical FDRs.** Stratified *q*-value thresholds outperform *E*-values in all tests (**Figure 3, Figure S2**). While stratified lFDR thresholds are superior to *E*-values in all tests, on most tests they surprisingly do not outperform *q*-values; this is in contrast to what we expect based on our theoretical results. We hypothesize that lFDR estimates are less robust than *q*-values to errors in *p*-values; these errors most likely arise because our empirical tests differ too much from the standard null model. We note that the Standard Pfam is not evaluated using ContextC and ClanOv (**Figure 3**), as these tests are based on the Pfam clans and context definitions, so the standard Pfam has a zero empirical FDR by construction in both. However, *q*-values outperform the Standard Pfam in two of the three fair tests (OrthoC and MarkovR) and *q*-values and the Standard Pfam perform similarly in the RevSeq test (**Figure S2**). Coherent with our theoretical results, the same trends were observed when the combined empirical *E*-values are controlled rather than the combined empirical FDRs (**Figure S7, Supplementary Methods**).

***Q*-value predictions are more informative than those of Standard Pfam and dPUC.** We desired to measure improvements not only of domain counts, which may be inflated for families with many small repeating units, but also of unique family counts. Further, we also measured performance with respect to a metric of information content that we developed based on the GO terms associated with different domain predictions (**Supplementary Methods**). To choose thresholds that are comparable to the amount of noise that Pfam currently permits, we calculated *p*- and *q*-value equivalents to the Standard Pam thresholds for each family (**Supplemental Results**). We used the median values of these distributions, which equal a stratified *q*-value threshold of 4e-4, and for *E*-values a *p*-value threshold of 1.3e-8 (**Supplemental Results**). We measure consistent improvements for q-values relative to the Standard Pfam using all metrics (between 4-7%, **Figure 4**). *E*-values predict 2.0% fewer domains than the Standard Pfam, but slightly outperform Pfam when considering the other metrics (**Figure 4**).

We also evaluated dPUC, a prediction method based on domain context [48,52]. dPUC also improves upon the Standard Pfam in all cases (**Figure 4**). While dPUC exhibits greater increases in domains than *q*-values, the two approaches are comparable at the family and amino acid coverage, and *q*-values have much higher protein and GO information content increases. This is because dPUC tends to predict more repeat domains (of the same family) and tends to restrict new predictions to proteins that already had Standard Pfam predictions. In contrast, *q*-values increase domains at the same rate as it increases proteins that did not have Standard Pfam predictions, which increase information the most. Thus, while stratified *q*-values predict fewer domains than dPUC, those domains tend to be more informative than the dPUC predictions at comparable FDRs.

**Empirical FDRs and *q*-values disagree, largely due to a small number of domain families.** The *q*-value thresholds and empirical FDRs were compared directly for agreement. We found large disagreements between *q*-values and our empirical FDRs tests (except for MarkovR; **Figure 5, Figure S8**). This again suggests that the *p*-values are problematic. Interestingly, the disagreement is proportionally larger for smaller FDRs, and it gets proportionally smaller as the FDR grows (**Figure 5**). We hypothesized that the problem might be due to a few domain families that are very noisy at stringent thresholds, and this subset becomes proportionally smaller as all families are allowed greater noise.

To further understand where the *q*-value and empirical FDR disagreements arise, we computed empirical FDRs separately per family using a threshold of *q*<=1e-2 (**Methods**). Note that this threshold corresponds to a higher level of noise than is typically obtained via the Standard Pfam (**Supplemental Results**) and this is desirable in this context as many domain families have very few predictions at strict thresholds. For these families, large deviations between the empirical FDRs and *q*-values may arise due to this low sampling; to address this, we modeled this random sampling to assess significance (**Methods**). We find that most families (92-99%, **Table S1**) have FDRs that are either close to their expected values or have differences that are not statistically significant (blue and black data in **Figure 6**, **Figure S9**).

**Empirical FDRs elevated in families with repetitive patterns.** Four of our tests (ClanOv, ContextC, OrthoC, and RevSeq) detected a notable number of families with significantly larger FDRs than expected (3-8%, **Table S1**). We found that these families are significantly enriched for families that contain coiled-coils, transmembrane domains, and low-complexity regions (**Figure 7; Methods**). There were fewer families with significantly smaller FDRs than expected (0-2%, **Table S1**), and we found no common pattern for these families. Only the MarkovR test conforms to our expectation, with no families having significantly larger FDRs than expected and 0.1% of families having significantly smaller FDRs than expected.

**Assigning domain families to noise classes.** To further investigate the properties of families with large and small FDR deviations, we used the four tests (excluding MarkovR) to assign families into mutually-exclusive classes by majority rule. The "increased-noise" families have significantly large positive deviations (see **Methods**; red in **Figure 6**) in at least three tests. The "decreased-noise" families have significantly large negative deviations (green in **Figure 6**) in at least three tests. Lastly, the families with "as-expected-noise" have small deviations (blue and some black in **Figure 6**) in at least three tests. There are 327 increased-noise families (2.7% of Pfam), only one decreased-noise family (HemolysinCabind), and 4433 families with as-expected noise (36%). The rest of the families in Pfam were not classified (7512 families, or 61% of Pfam). The complete lists of increased noise and as-expected noise families are available in the supplementary materials (**Supplementary Files 1 and 2**).

**Pfam threshold curation somewhat compensates for underestimated noise**. Here we investigate if the curation of thresholds that Pfam performs compensates for underestimated FDRs. We find that the equivalent UniRef50 $q$-values of the Standard Pfam curated thresholds are usually smaller for the increased-noise families than the as-expected-noise families, with the median increased-noise family $q$-value 19 times smaller than the median as-expected-noise family $q$-value. However, there is no clear separation between both $q$-value distributions (**Figure S10**). In particular, 15% of the increased-noise families have a $q$-value threshold larger than the median as-expected-noise threshold. Overall, while Pfam has more stringent thresholds for increased-noise families than for families with as-expected noise, many increased-noise family thresholds remain more permissive than they should be.

**The lFDR outperforms $q$-values in families with as-expected noise.** Empirical FDRs agree more with the stratified $q$-values in the families with as-expected noise than in all families combined, although some disagreement remains (**Figure S11**). In this subset of families, lFDRs outperforms $q$-values in all tests (**Figure 8**), as we expect from our theoretical results when the underlying $p$-values are correct. Compared to the Standard Pfam, the domain count improvement at a $q$-value threshold of 4e-4 increases from 6.7% in all families to 8.8% in the subset of families with as-expected noise (similar increases are observed on all metrics, see **Figure S12**), and lFDRs further improve upon $q$-values. This is indicative of the potential broader usefulness of stratified lFDR estimates should $p$-values of all families improve in the future.

**Tiered stratified $q$-values.** The previous methods describe a single threshold set on the domain $p$-values via the stratified $q$-value or lFDR analysis. However, HMMER provides additional information in the form of "sequence" *p*-values, which score the presence of domain families as the combined evidence of all of its domain repeats. Here we are interested in setting thresholds on both kinds of *p*-values simultaneously. Only 2.3% of Pfam families have non-redundant sequence and domain Standard Pfam thresholds [14]. We can instead define these two-tier thresholds in terms of the FDR. In the first tier, we compute *q*-values from the sequence *p*-values and set the threshold $Q_{seq}$. In the second tier, we have a pruned list of domain *p*-values left, for the domains in sequences that satisfied the sequence threshold only, on which we compute *q*-values again (which correspond to a FDR conditional on the first FDR filter) and set the second threshold $Q_{dom|seq}$. We show in the **Supplementary Methods** that the final FDR is approximately $Q_{seq}+Q_{dom|seq}$ when both thresholds are small and under a simplifying assumption of independence. For simplicity, we use $Q_{seq}=Q_{dom|seq}$ in our tests. While the same could be done with lFDRs, we decided to use *q*-values only since they are more robust.

Tiered *q*-values predict many more domains, at any fixed empirical FDR, than domain *q*-values and domain lFDRs, our previous two best statistics, consistently and by very large margins (**Figure S2**). We expected tiered *q*-values to predict more domains because the procedure should easily predict additional repeating domains when there is strong evidence for a family to be present. Remarkably, tiered *q*-values outperform other methods in predicting new families per sequence (**Figure S3**), which shows that this approach is predicting some families that are otherwise not predicted at all (that is, the entire signal of these families comes from the combined power of repeating units, none of which is significant by itself). There is also a very large increase in amino acid coverage (**Figure S4**), and a smaller increase in protein coverage (**Figure S5**) and GO information content (**Figure S6**). However, our theoretical FDR estimates for the tiered *q*-value procedure are more inaccurate than for domain *q*-values (**Figure S8**), and this additional inaccuracy remains in the subset of families with as-expected noise (**Figure S11**). For this reason we treat tiered stratified *q*-values, which are more powerful than domain-only *q*-values, as an experimental approach, as work remains to be done in wielding them to accurately control the FDR.

Our tiered stratified *q*-value approach also compares favorably to dPUC [48]. Tiered *q*-values match the superior domain improvements of dPUC, slightly improve upon dPUC and domain stratified *q*-values in family and amino acid coverage, and outperform dPUC at protein coverage and GO information content (**Figure S13**). Thus, tiered *q*-values retain all the strengths of domain *q*-values while powerfully leveraging the limited context information of repeating domains present in sequence *q*-values.

**DISCUSSION**

In multiple hypothesis testing, the FDR and lFDR are quantities that allow for simple balancing of the proportion of false positives and the posterior error probability. The *q*-value is a popular statistic for controlling the FDR that is both conservative and more powerful than previous FDR procedures such as the one from Benjamini and Hochberg [22]. Benchmarks based on empirical FDRs have been a part of many recent works studying protein and DNA homology [47,48,52,53], although they are based on expensive simulations rather than estimating FDRs directly from *p*-values (or *E*-values), as *q*-values do very efficiently. Our work is, to the best of our knowledge, the first attempt at applying *q*-values and lFDRs to domain identification, thus advancing the statistics of this field.

Domain prediction is one case where stratified FDR and lFDR control are desirable, since domain families occur with vastly different frequencies and are thus associated with differing amounts of true signal. However, the same can be said of many other applications, such as BLAST-based orthology prediction [20], where some proteins may have orders of magnitude more true orthologs than others. It is also possible FDR and lFDR control will improve iterative profile database searches, such as PSI-BLAST [6], as well as numerous other sequence analysis problems.

Our theoretical work revealed that the lFDR, which is the Bayesian posterior probability that a prediction is false, is the optimal quantity to control in stratified problems. Stratified lFDR control has previously been found to optimize stratified thresholds in the related problem of minimizing the combined false non-discovery rate while controlling the combined FDR [37]. The lFDR also arises naturally in Bayesian classification problems [43]. Stratified lFDR thresholds are appealing since they ensure the worst predictions of each stratum have the same posterior probability of being false. However, we found that estimated *q*-values are more robust than our lFDR estimates when the *p*-values are imperfect [44] (**Figure 3**).

We extended the domain stratified *q*-value approach into what we call tiered stratified *q*-values, by setting *q*-value thresholds on both of the sequence and domain statistics reported by HMMER. While accurately determining the final FDR of this procedure remains a challenge, we show that tiered *q*-values successfully exploit the signal contained in repeating domain units of the same family to produce more domain predictions (**Figure S13**). There are other successful approaches, such as dPUC [48] and CODD [52], that exploit the broader concept of domain context (or co-occurrence) to improve domain predictions. It is remarkable that tiered *q*-values perform as well as or better than these context-based methods under all metrics (**Figure S13**) although tiered *q*-values only utilize the context signal of repeating domains, while dPUC additionally utilizes the context signal present between domains of different families [48]. Therefore, tiered stratified *q*-values could be combined in the future with the inter-family context information that the dPUC framework employs to yield further improvements in domain prediction.

In order to test our approach for domain prediction, we introduced a suite of empirical FDR tests. We evaluate these tests in the **Supplementary Results**, and together we find them to be a powerful means for testing the correctness of predictions. Four of our tests consistently revealed flaws with the underlying statistics of domain prediction. We found a very strong association between significantly underestimated FDRs and the presence of coiled coils, transmembrane domains, and other low-complexity regions. The problems that these categories of domain families pose have been noted separately elsewhere [46,54,55], and *ad hoc* solutions to these problems have been proposed before [54,56]. However, none of these solutions are implemented by standard sequence similarity software such as BLAST and HMMER [56]; in our view, obtaining correct statistics for families with repetitive patterns that do not imply homology should be the top priority of the field of sequence homology. Nevertheless, our benchmarks indicate that most families in Pfam produce correct statistics (as-expected noise), and for these families, the advantage of using *q*-value and lFDR statistics is clear. In the future, the standard sequence similarity software packages should be able to compute and report these statistics natively rather than as a post-processing step as is done here.

The basis of our work is a general theorem that shows how to maximize predictions while controlling for noise and taking advantage of having naturally stratified statistical tests. Whether the combined FDR or *E*-value is constrained, the unifying principle is that equal stratified lFDR thresholds are required to maximize predictions. Besides limits on sample size, the strata may be arbitrary, which allows us to utilize our result in many broad forms of multiple hypothesis testing. For example, if one has *p*-values across multiple microarrays, each of which may contain different sets of genes from different organisms, we will get the most positives while controlling the FDR across the entire dataset by finding the per-microarray lFDR threshold that gives the desired combined FDR. A more similar example is that of motif scanning, for example *in silico* transcription factor (TF) binding site identification, where the position weight matrix of each TF is used to predict binding, which may yield a *p*-value per match [57], and the number of binding sites per TF may vary by orders of magnitude across different TFs. Here, as in the protein domain case, it may be beneficial to compute lFDRs per TF and set equal lFDR thresholds across TFs. In the case of protein domains, the next logical step is to further stratify *p*-values by taxonomy of the protein database in addition to domain family, since it is well known that domain abundance varies greatly across the main kingdoms of life (archaea, bacteria, eukarya, and viruses) [58,59]. In sum, while we have demonstrated the practical utility of our theoretical contributions to domain prediction, they are additionally relevant for a wide range of applications in bioinformatics and beyond.

**METHODS**

**HMMER *p*-values.** A *p*-value distribution is required to estimate *q*-values and lFDRs. HMMER reports two kinds of *p*-values. The "sequence" *p*-value encompasses every domain of the same family on a protein sequence, while the "domain" *p*-value is limited to each domain instance. The sequence *p*-value is thus reporting on whether the protein sequence as a whole contains similarity to the HMM, whereas the domain *p*-value scores each individual domain unit found within the sequence. We obtained domain predictions with *p*-values on UniRef50 [35] and OrthoMCL5 [20] proteins using hmmsearch from HMMER 3.0 and HMMs from Pfam 25 with the following options: the heuristic filters "--F1 1e-1 --F2 1e-1 --F3 1e-2" allow sequence predictions with "stage 1/2/3" *p*-values of 0.1, 0.1, and 0.01 instead of the defaults of 0.02, 1e-3, and 1e-5 respectively. Moreover, we force *p*-values by setting "-Z 1 --domZ 1". Lastly, we remove domains with *p*>0.01 by adding "-E 1e-2 --domE 1e-2".

**Overview of *q*-value and lFDR estimation for domains.** For each domain family HMM, we use its HMMER *p*-values over a protein database to estimate *q*-values and lFDRs. We use standard methods [27,44] adapted for censored tests since HMMER3 only reports the most significant *p*-values (standard methods require all *p*-values). Notably, HMMER3 does not provide complete *p*-value sets even if filters are removed [60], and only small *p*-values are accurate [19], so the full set of *p*-values is not useful. Moreover, it is desirable to keep the filters to reduce HMMER3's runtime. The **Supplementary Methods** reviews these standard methods and details our adaptations for domains. Briefly, we remove overlaps within the same sequence between domain predictions, ranking by *p*-value, before computing *q*-values and lFDRs; otherwise, the amount of true positive may be overestimated because overlapping related domains will be double counted, a situation frequently encountered within some Pfam clans. Secondly, the standard *q*-value and lFDR approaches require all *p*-values solely to estimate $\pi_0$, here roughly the proportion of proteins that do not contain an instance of a given domain family. We set this parameter to 1, which we expect to result in slightly more conservative *q*-values and lFDR estimates than we would have otherwise. Our software for computing stratified *q*-values, lFDR estimates and tiered *q*-values from HMMER3 is DomStratStats 1.03, available at http://viiia.org/domStratStats/?l=en-us.

**Baseline threshold methods.** We wish to compare domains predicted with $q$-values and lFDRs to standard domain prediction approaches, over a range of relevant empirical FDRs. We vary thresholds based on stratified $q$-values and lFDRs, and compare their performances to thresholds varied by $E$-values and extensions of the Standard Pfam thresholds. Stratified domain $E$-values are computed from the HMMER $p$-values by multiplying them by the number of proteins in UniRef50, as hmmsearch would compute it. We note that our benchmarks are unaffected by transformations of the test statistic that preserve rankings. For this reason, $p$-values and unstratified $q$-values (of the complete set of $p$-values) also perform the same as stratified $E$-values, since these transformations preserve rankings. On the other hand, stratified $q$-values and stratified lFDRs change domain rankings between strata. We also compare to the "Standard Pfam," a more complex thresholding system that consists of two expert-curated thresholds per family, one for the domain and sequence bitscores respectively (in the Pfam documentation, these are called "gathering" thresholds) [14]. Since all domain bitscores are provided by HMMER, we can extend these curated thresholds (which together produce a single datapoint with a fixed FDR) by shifting all fixed bitscore thresholds by constant amounts as described previously [48], allowing us to explore a range of FDRs. In all cases, overlaps within sequences are removed by $p$-value (equivalent to $E$-value) ranking. Overlaps between families in the "nesting" list (i.e., those observed in UniRef with Standard Pfam thresholds) are not removed (**Supplementary Methods**). All methods use a permissive overlap definition [61] (**Supplementary Methods**), except for the Standard Pfam (there overlaps of even one amino acid are removed [14]). For each domain family, the Standard Pfam thresholds are mapped to $p$-values, $q$-values, and lFDRs, and the medians of these distributions are used in many plots (**Figure S14; Supplementary Results**).

**Empirical FDR tests.** A crucial component of our work is the introduction of a suite of tests designed to measure empirical FDRs using biologically-motivated definitions of TPs and FPs. The "standard" biological sequence null model, which most software from BLAST to HMMER use, defines a random sequence as a sequence of independent and identically distributed amino acids, which are derived from the background distribution. Domains predicted on these random sequences produce a distribution of random bit scores from which $p$-values are computed. The five empirical tests we use instead label every prediction as either a TP or a FP, and these labels are used to compute empirical FDRs and $E$-values (the latter referring to the number of type I errors, or FPs). Each test makes different assumptions in assigning TPs and FPs, and thus together they provide independent and complementary evaluations. We describe our two primary tests in detail next, while the other three tests are described in the **Supplementary Methods**.

**a) Clan Overlap (ClanOv).** This test is inspired by [46]. After domains are predicted on each sequence and ranked by *p*-value, only overlaps between domains of the same clan are eliminated. Each remaining domain is labeled a FP if it overlaps a higher-ranking domain of a different clan, and otherwise it is labeled a TP. Domains are considered to overlap using the permissive overlap definition (**Supplementary Methods**). This test does not evaluate the Standard Pfam fairly because it is assigned an FDR of zero, partly because the Standard Pfam thresholds are directly optimized on a similar test to prevent inter-clan overlaps [14], but also because our "nesting" list of allowed overlaps is defined using the Standard Pfam (**Supplementary Methods**). Overlaps are removed before reporting the number of domain predictions (used in the *y*-axis of plots such as those in **Figure 3**). Since *q*-values and lFDRs are computed on domains without overlaps, but ClanOv requires overlaps to measure empirical FDRs, here domains that overlap higher-ranking domains must be preserved and must have *q*-values and lFDRs. We assign these *q*-values and lFDRs by interpolation.

**b) Context Coherence (ContextC).** This test is an extension of one proposed in [47], who first suggested that co-occurring domains could be used to estimate the FDR. Here, given a predefined list $L$ of context family pairs (families that have been found to co-occur within the same sequence) and domains ranked by *p*-value, a domain is labeled as a TP if it is the highest-ranking domain or a higher-ranking domain can be found such that their family pair is in $L$; otherwise the domain is labeled a FP. This test does not evaluate the Standard Pfam thresholds fairly because they are assigned an FDR of zero, since context family pairs are defined by the Standard Pfam observations (**Supplementary Methods**).

**Computing empirical FDRs.** Using any of the above empirical tests to label domain predictions as either TPs or FPs, we evaluate empirical FDRs at two levels. Briefly, the "method-level" FDR is an evaluation of an entire scoring method (*q*-values, *E*-values, etc.) when all domain families are pooled, whereas the "family-level" FDR evaluates the accuracy of *q*-values separately per family. These quantities are biased estimators of the corresponding true pFDRs, but as variants of the mFDR, are asymptotically unbiased under weak dependence [45]. For a threshold *t*, let $\text{TP}_{ij}(t)$ and $\text{FP}_{ij}(t)$ be the observed number of true positives and false positives, respectively, for domain family *j* in protein sequence *i*.

**a) Method-level FDR and *E*-value.** Given empirical TP, FP labels as described above, we compute FDRs and *E*-values across domain families. First, the empirical protein-level FDR, or epFDR, of protein *i* with domain predictions is computed by summing over all domain families *j* as

$$\text{epFDR}_i(t) = \frac{\sum_j \text{FP}_{ij}(t)}{\sum_j \text{TP}_{ij}(t) + \text{FP}_{ij}(t)} .$$

Then the method-level empirical FDR is the mean $\text{epFDR}_i(t)$ over all proteins $i$ with predictions, which corresponds to the expected FDR per protein. This per-protein FDR normalizes for proteins having varying numbers of domain predictions, especially since certain proteins carry hundreds of domain instances. The standard error used in plots is computed from the $\text{epFDR}_i(t)$ samples. The empirical $E$-value of protein $i$ is $\sum_j \text{FP}_{ij}(t)$, used for error bars, and the final empirical $E$-value is the sum of the per-protein $E$-values; these empirical $E$-values therefore measure the number of false positives we expect in total, across proteins and domain families.

    **b) Domain family-level FDR.** This procedure measures, for each domain family $j$ and an empirical test, the deviation between the empirical FDRs and the $q$-value threshold; ideally these measures agree. We measured all deviations at $q=1\text{e-}2$. The empirical family-level FDR, or efFDR, of this family $j$ is defined as

$$\text{efFDR}_j(q) = \frac{1 + \sum_i \text{FP}_{ij}(q)}{1 + \sum_i \text{TP}_{ij}(q) + \text{FP}_{ij}(q)} ,$$

where true and false positives are summed across proteins $i$, and the log-deviation is defined as

$$\text{LD}_j = \log_2 \frac{\text{efFDR}_j(q)}{q} .$$

A pseudocount of 1 is used in $\text{efFDR}_j(q)$ so $\text{LD}_j$ is always defined (that is, $\text{efFDR}_j(q)>0$ even if $\sum_i \text{FP}_{ij}(q)=0$). Since most families have very few predictions, the LD may be artificially large or small. We test for statistical significance by computing a two-tailed $p$-value ($p_{Poisson}$) of the empirical $E$-value $\sum_i \text{FP}_{ij}(q)$ using the Poisson distribution with parameter $q\left(\sum_i \text{TP}_{ij}(q) + \text{FP}_{ij}(q)\right)$, which is the expected number of FPs given the number of observations. This statistical evaluation does not use the pseudocount. The $p_{Poisson}$ distribution is used to compute $q$-values ($q_{Poisson}$, unrelated to the $q$-value threshold set on the domains), and a measurement is called significant if $q_{Poisson} <= 1\text{e-}3$. In addition, families are separated into three groups by effect size: positive deviations if $\text{LD}_j > 2$, negative deviations if $\text{LD}_j < -2$, and small deviations if $|\text{LD}_j| <= 2$.

**Domain Prediction Using Context (dPUC) method.** We also benchmarked dPUC 2.0, a method that improves domain prediction by taking into account the "context," or the presence of other domain predictions [48]. dPUC 2.0 now works with HMMER3, among other improvements that will be described elsewhere. Context family pair counts were derived from Pfam 25 on UniProt proteins. The "candidate domain *p*-value threshold" of dPUC can be tuned to achieve a desired FDR. Here we marked *p*<=1e-4 in plots, which gives comparable empirical FDRs to *q*<=4e-4 on MarkovR and OrthoC. dPUC is not evaluated in ContextC because both are based on domain context (dPUC would have a zero empirical FDR), nor in ClanOv because dPUC requires overlap removal while ClanOv requires observing overlaps to compute its FDR. The dPUC 2.0 software is available at http://compbio.cs.princeton.edu/dpuc.

**Categorizing families with repetitive patterns.** The programs PairCoil2 [62], TMHMM [63], and SEG [64], were run on UniRef50 using standard parameters to predict coiled coils, transmembrane domains, and low-complexity regions, respectively. Each Pfam family observed at least 4 times in UniRef50 was associated with a category if it overlapped its predictions more than half the time this family was observed. If a family was associated with multiple categories, average coverage of the domain instances was used to assign it to the category with the strongest association. All unassigned families were categorized as "other".

## DATA AVAILABILITY STATEMENT
Our research is based entirely on the following public data: the Pfam domain database, the UniProt protein sequence database, the OrthoMCL orthology prediction database, and the Gene Ontology database. The only exception is the 2$^{nd}$ order Markov random UniRef50-based protein sequence dataset, which is available at http://viiia.org/randProt/?l=en-us. Any intermediate and resulting data not already provided in the links in the methods or as supplementary files is available upon request.

## ACKNOWLEDGEMENTS
We thank our reviewers, as well as all members of the Singh, Llinás, and Storey groups for helpful discussions about this work. This work was supported by the National Science Foundation [Graduate Research Fellowship DGE 0646086 to AO]; and the National Institutes of Health [1 R21-AI085415 to MS and ML, Center of Excellence P50 GM071508 to the Lewis-Sigler Institute].

## REFERENCES

1. Needleman SB, Wunsch CD. A general method applicable to the search for similarities in the amino acid sequence of two proteins. Journal of Molecular Biology. 1970;48: 443–453. doi:10.1016/0022-2836(70)90057-4
2. Smith TF, Waterman MS. Identification of common molecular subsequences. Journal of Molecular Biology. 1981;147: 195–197. doi:10.1016/0022-2836(81)90087-5
3. Lipman DJ, Pearson WR. Rapid and sensitive protein similarity searches. Science. 1985;227: 1435–1441. doi:10.1126/science.2983426
4. Karlin S, Altschul SF. Methods for assessing the statistical significance of molecular sequence features by using general scoring schemes. P Natl Acad Sci U S A. 1990;87: 2264–2268.
5. Altschul SF, Gish W, Miller W, Myers EW, Lipman DJ. Basic local alignment search tool. J Mol Biol. 1990;215: 403–10. doi:10.1006/jmbi.1990.9999
6. Altschul SF, Madden TL, Schäffer AA, Zhang J, Zhang Z, Miller W, et al. Gapped BLAST and PSI-BLAST: a new generation of protein database search programs. Nucleic Acids Res. 1997;25: 3389–3402.
7. Eddy SR. Profile hidden Markov models. Bioinformatics. 1998;14: 755–763. doi:10.1093/bioinformatics/14.9.755
8. Barrett C, Hughey R, Karplus K. Scoring hidden Markov models. Comput Appl Biosci. 1997;13: 191–199. doi:10.1093/bioinformatics/13.2.191
9. Madera M. Profile Comparer: a program for scoring and aligning profile hidden Markov models. Bioinformatics. 2008;24: 2630–2631. doi:10.1093/bioinformatics/btn504
10. Söding J. Protein homology detection by HMM-HMM comparison. Bioinformatics. 2005;21: 951–960. doi:10.1093/bioinformatics/bti125
11. Sadreyev R, Grishin N. COMPASS: A Tool for Comparison of Multiple Protein Alignments with Assessment of Statistical Significance. Journal of Molecular Biology. 2003;326: 317–336. doi:10.1016/S0022-2836(02)01371-2
12. Altschul SF, Wootton JC, Zaslavsky E, Yu Y-K. The Construction and Use of Log-Odds Substitution Scores for Multiple Sequence Alignment. PLoS Comput Biol. 2010;6: e1000852. doi:10.1371/journal.pcbi.1000852
13. Haussler D, Krogh A, Mian IS, Sjolander K. Protein modeling using hidden Markov models: analysis of globins. System Sciences, 1993, Proceeding of the Twenty-Sixth Hawaii International Conference on. 1993. pp. 792–802 vol.1. doi:10.1109/HICSS.1993.270611
14. Punta M, Coggill PC, Eberhardt RY, Mistry J, Tate J, Boursnell C, et al. The Pfam protein families database. Nucleic Acids Research. 2011;40: D290–D301. doi:10.1093/nar/gkr1065
15. Krogh A, Brown M, Mian IS, Sjölander K, Haussler D. Hidden Markov Models in Computational Biology : Applications to Protein Modeling. J Mol Biol. 1994;235: 1501–1531. doi:10.1006/jmbi.1994.1104
16. Wilson D, Pethica R, Zhou Y, Talbot C, Vogel C, Madera M, et al. SUPERFAMILY--sophisticated comparative genomics, data mining, visualization and phylogeny. Nucl Acids Res. 2009;37: D380–386. doi:10.1093/nar/gkn762
17. Letunic I, Doerks T, Bork P. SMART 6: recent updates and new developments. Nucl Acids Res. 2009;37: D229–232. doi:10.1093/nar/gkn808
18. Eddy SR. Accelerated Profile HMM Searches. PLoS Comput Biol. 2011;7: e1002195. doi:10.1371/journal.pcbi.1002195
19. Eddy SR. A Probabilistic Model of Local Sequence Alignment That Simplifies Statistical Significance Estimation. PLoS Comput Biol. 2008;4: e1000069. doi:10.1371/journal.pcbi.1000069
20. Chen F, Mackey AJ, Jr CJS, Roos DS. OrthoMCL-DB: querying a comprehensive multi-species collection of ortholog groups. Nucleic Acids Res. 2006;34: D363–D368. doi:10.1093/nar/gkj123
21. Sonnhammer ELL, Eddy SR, Durbin R. Pfam: A comprehensive database of protein domain

21. families based on seed alignments. Proteins: Structure, Function, and Genetics. 1997;28: 405–420. doi:10.1002/(SICI)1097-0134(199707)28:3<405::AID-PROT10>3.0.CO;2-L
22. Benjamini Y, Hochberg Y. Controlling the False Discovery Rate: A Practical and Powerful Approach to Multiple Testing. Journal of the Royal Statistical Society Series B (Methodological). 1995;57: 289–300.
23. Storey JD. A direct approach to false discovery rates. Journal of the Royal Statistical Society: Series B (Statistical Methodology). 2002;64: 479–498. doi:10.1111/1467-9868.00346
24. Efron B, Tibshirani R, Storey JD, Tusher V. Empirical Bayes Analysis of a Microarray Experiment. Journal of the American Statistical Association. 2001;96: 1151–1160. doi:10.1198/016214501753382129
25. Glivenko V. Sulla determinazione empirica della legge di probabilita. Giorn Ist Ital Attuari. 1933;4: 92–99.
26. Cantelli FP. Sulla determinazione empirica delle leggi di probabilita. Giorn Ist Ital Attuari. 1933;4: 221–424.
27. Storey JD, Tibshirani R. Statistical significance for genomewide studies. Proceedings of the National Academy of Sciences of the United States of America. 2003;100: 9440–9445. doi:10.1073/pnas.1530509100
28. Sun W, Cai TT. Oracle and Adaptive Compound Decision Rules for False Discovery Rate Control. Journal of the American Statistical Association. 2007;102: 901–912. doi:10.1198/016214507000000545
29. Schork AJ, Thompson WK, Pham P, Torkamani A, Roddey JC, Sullivan PF, et al. All SNPs Are Not Created Equal: Genome-Wide Association Studies Reveal a Consistent Pattern of Enrichment among Functionally Annotated SNPs. PLoS Genet. 2013;9: e1003449. doi:10.1371/journal.pgen.1003449
30. Keller A, Nesvizhskii AI, Kolker E, Aebersold R. Empirical statistical model to estimate the accuracy of peptide identifications made by MS/MS and database search. Anal Chem. 2002;74: 5383–5392.
31. Käll L, Storey JD, MacCoss MJ, Noble WS. Assigning Significance to Peptides Identified by Tandem Mass Spectrometry Using Decoy Databases. J Proteome Res. 2008;7: 29–34. doi:10.1021/pr700600n
32. Käll L, Storey JD, MacCoss MJ, Noble WS. Posterior error probabilities and false discovery rates: two sides of the same coin. J Proteome Res. 2008;7: 40–44. doi:10.1021/pr700739d
33. Choi H, Nesvizhskii AI. Semisupervised model-based validation of peptide identifications in mass spectrometry-based proteomics. J Proteome Res. 2008;7: 254–265. doi:10.1021/pr070542g
34. Choi H, Ghosh D, Nesvizhskii AI. Statistical validation of peptide identifications in large-scale proteomics using the target-decoy database search strategy and flexible mixture modeling. J Proteome Res. 2008;7: 286–292. doi:10.1021/pr7006818
35. The UniProt Consortium. Reorganizing the protein space at the Universal Protein Resource (UniProt). Nucleic Acids Research. 2011; doi:10.1093/nar/gkr981
36. Efron B. Simultaneous inference: When should hypothesis testing problems be combined? Ann Appl Stat. 2008;2: 197–223. doi:10.1214/07-AOAS141
37. Cai TT, Sun W. Simultaneous Testing of Grouped Hypotheses: Finding Needles in Multiple Haystacks. Journal of the American Statistical Association. 2009;104: 1467–1481. doi:10.1198/jasa.2009.tm08415
38. Hu JX, Zhao H, Zhou HH. False Discovery Rate Control With Groups. J Am Stat Assoc. 2010;105: 1215–1227. doi:10.1198/jasa.2010.tm09329
39. Zou J, Hong G, Zheng J, Hao C, Wang J, Guo Z. Evaluating FDR and stratified FDR control approaches for high-throughput biological studies. 2012 IEEE Symposium on Robotics and Applications (ISRA). 2012. pp. 684–686. doi:10.1109/ISRA.2012.6219282

40. Sun L, Craiu RV, Paterson AD, Bull SB. Stratified false discovery control for large-scale hypothesis testing with application to genome-wide association studies. Genetic Epidemiology. 2006;30: 519–530. doi:10.1002/gepi.20164
41. Li C, Li M, Lange EM, Watanabe RM. Prioritized Subset Analysis: Improving Power in Genome-wide Association Studies. Hum Hered. 2007;65: 129–141. doi:10.1159/000109730
42. Storey JD. False Discovery Rate. In: Lovric M, editor. International Encyclopedia of Statistical Science. Springer Berlin Heidelberg; 2014. pp. 504–508. Available: http://link.springer.com/referenceworkentry/10.1007/978-3-642-04898-2_248
43. Storey JD. The positive false discovery rate: a Bayesian interpretation and the q-value. Ann Statist. 2003;31: 2013–2035. doi:10.1214/aos/1074290335
44. Strimmer K. A unified approach to false discovery rate estimation. BMC Bioinformatics. 2008;9: 303. doi:10.1186/1471-2105-9-303
45. Storey JD, Taylor JE, Siegmund D. Strong control, conservative point estimation and simultaneous conservative consistency of false discovery rates: a unified approach. Journal of the Royal Statistical Society Series B. 2004;66: 187–205. doi:10.1111/j.1467-9868.2004.00439.x
46. Mistry J, Finn RD, Eddy SR, Bateman A, Punta M. Challenges in homology search: HMMER3 and convergent evolution of coiled-coil regions. Nucl Acids Res. 2013;41: e121–e121. doi:10.1093/nar/gkt263
47. Terrapon N, Gascuel O, Marechal E, Brehelin L. Fitting hidden Markov models of protein domains to a target species: application to Plasmodium falciparum. BMC Bioinformatics. 2012;13: 67. doi:10.1186/1471-2105-13-67
48. Ochoa A, Llinás M, Singh M. Using context to improve protein domain identification. BMC Bioinformatics. 2011;12: 90. doi:10.1186/1471-2105-12-90
49. Karplus K, Karchin R, Shackelford G, Hughey R. Calibrating E-values for hidden Markov models using reverse-sequence null models. Bioinformatics. 2005;21: 4107–4115. doi:10.1093/bioinformatics/bti629
50. Ashburner M, Ball CA, Blake JA, Botstein D, Butler H, Cherry JM, et al. Gene Ontology: tool for the unification of biology. Nat Genet. 2000;25: 25–29. doi:10.1038/75556
51. Forslund K, Sonnhammer ELL. Predicting protein function from domain content. Bioinformatics. 2008;24: 1681–1687. doi:10.1093/bioinformatics/btn312
52. Terrapon N, Gascuel O, Marechal E, Brehelin L. Detection of new protein domains using co-occurrence: application to Plasmodium falciparum. Bioinformatics. 2009;25: 3077–3083. doi:10.1093/bioinformatics/btp560
53. Wheeler TJ, Clements J, Eddy SR, Hubley R, Jones TA, Jurka J, et al. Dfam: a database of repetitive DNA based on profile hidden Markov models. Nucleic Acids Research. 2012;41: D70–D82. doi:10.1093/nar/gks1265
54. Rackham OJL, Madera M, Armstrong CT, Vincent TL, Woolfson DN, Gough J. The Evolution and Structure Prediction of Coiled Coils across All Genomes. Journal of Molecular Biology. 2010;403: 480–493. doi:10.1016/j.jmb.2010.08.032
55. Wong W-C, Maurer-Stroh S, Eisenhaber F. More Than 1,001 Problems with Protein Domain Databases: Transmembrane Regions, Signal Peptides and the Issue of Sequence Homology. PLoS Comput Biol. 2010;6: e1000867. doi:10.1371/journal.pcbi.1000867
56. Wong W-C, Maurer-Stroh S, Schneider G, Eisenhaber F. Transmembrane Helix: Simple or Complex. Nucl Acids Res. 2012; doi:10.1093/nar/gks379
57. Hartmann H, Guthöhrlein EW, Siebert M, Luehr S, Söding J. P-value-based regulatory motif discovery using positional weight matrices. Genome Res. 2013;23: 181–194. doi:10.1101/gr.139881.112
58. Apic G, Gough J, Teichmann SA. Domain combinations in archaeal, eubacterial and eukaryotic proteomes. Journal of Molecular Biology. 2001;310: 311–325. doi:10.1006/jmbi.2001.4776

59. Yang S, Bourne PE. The Evolutionary History of Protein Domains Viewed by Species Phylogeny. PLoS ONE. 2009;4: e8378. doi:10.1371/journal.pone.0008378
60. Eddy SR. HMMER3 is stubborn. In: Cryptogenomicon [Internet]. 19 Sep 2011 [cited 4 Jan 2013]. Available: http://selab.janelia.org/people/eddys/blog/?p=508
61. Yeats C, Redfern OC, Orengo C. A fast and automated solution for accurately resolving protein domain architectures. Bioinformatics. 2010;26: 745–751. doi:10.1093/bioinformatics/btq034
62. McDonnell AV, Jiang T, Keating AE, Berger B. Paircoil2: Improved Prediction of Coiled Coils from Sequence. Bioinformatics. 2006;22: 356–358. doi:10.1093/bioinformatics/bti797
63. Krogh A, Larsson B, von Heijne G, Sonnhammer ELL. Predicting transmembrane protein topology with a hidden markov model: application to complete genomes. Journal of Molecular Biology. 2001;305: 567–580. doi:10.1006/jmbi.2000.4315
64. Wootton JC. Non-globular domains in protein sequences: Automated segmentation using complexity measures. Computers & Chemistry. 1994;18: 269–285. doi:10.1016/0097-8485(94)85023-2

**FIGURES AND LEGENDS**

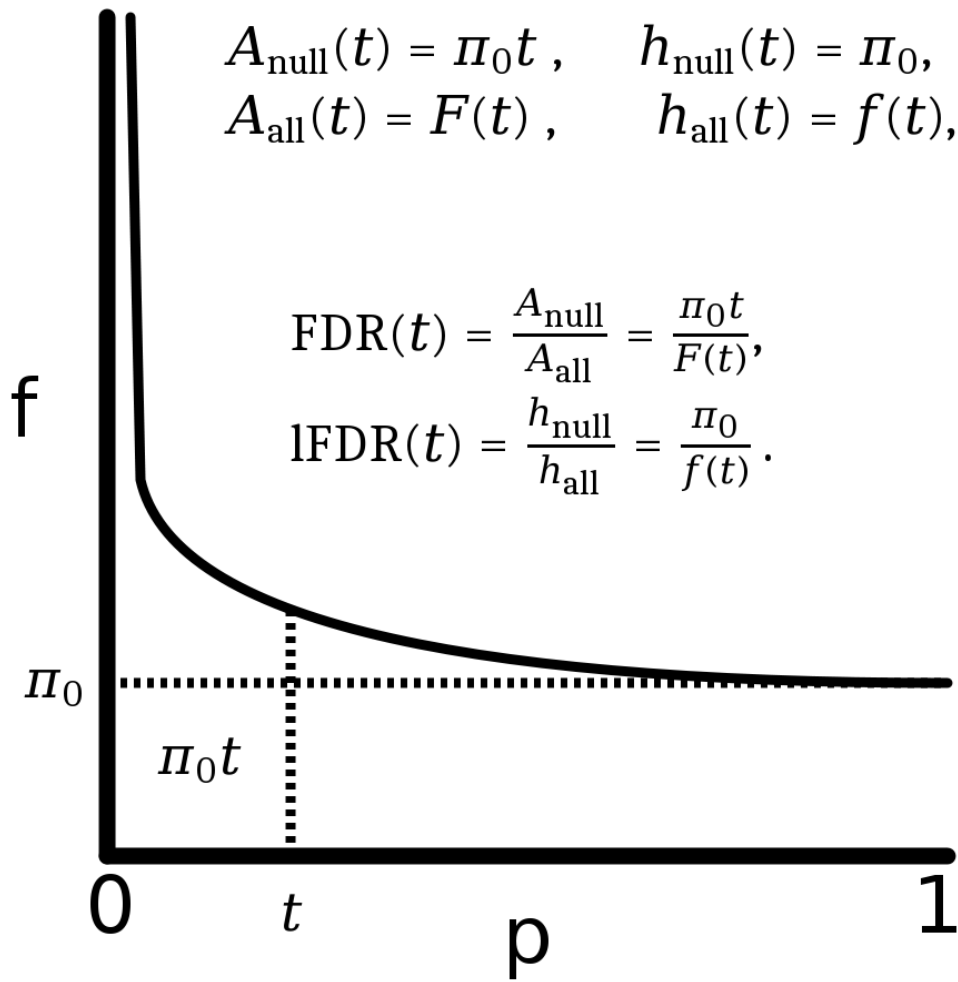

**Figure 1. The Bayesian FDR and lFDR.** Both quantities assume a distribution of *p*-values with two components: "null" *p*-values which are uniformly distributed (its height is $\pi_0 <= 1$), and "alternative" *p*-values which should peak at *p*=0. The area of the null component at a *p*-value threshold *t* is simply $\pi_0 t$, while the total area is the cumulative density function $F(t)$. The Bayesian FDR is the proportion of the area with $p <= t$ that corresponds to the null component. The lFDR parallels the Bayesian FDR but is a ratio of densities (heights) rather than areas.

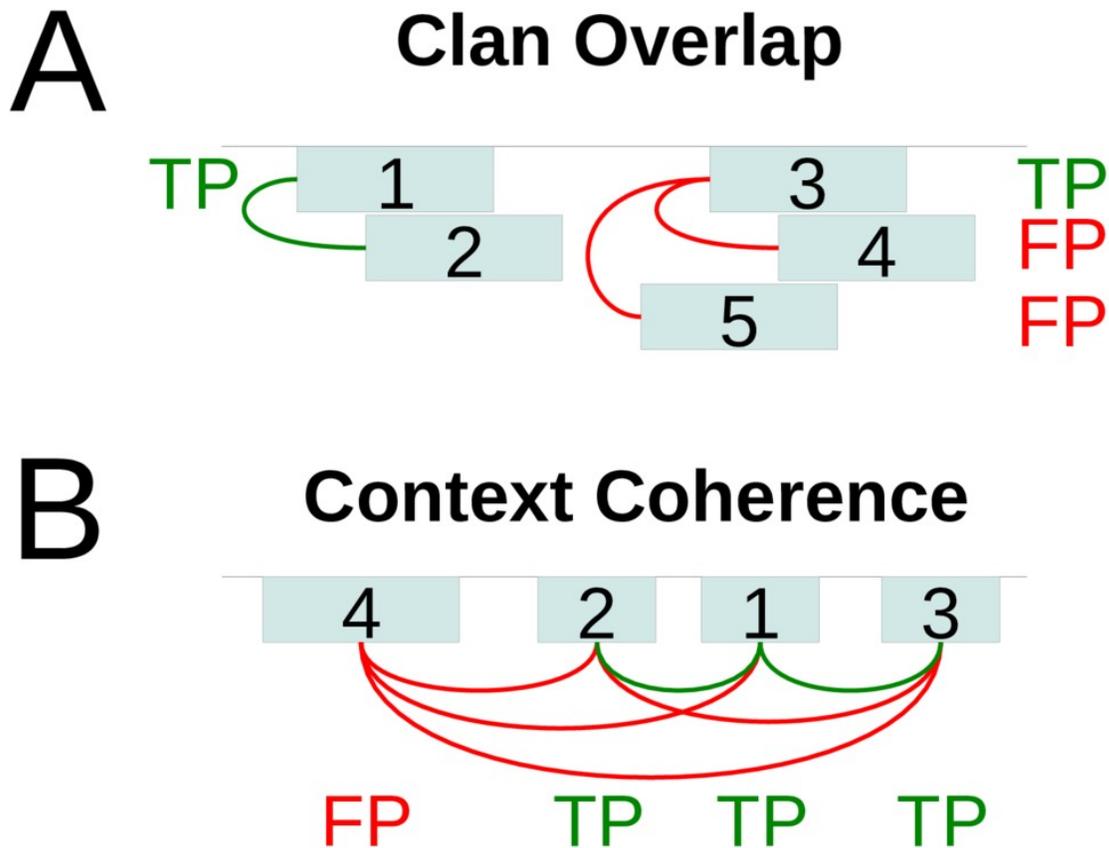

**Figure 2. Illustration of the empirical FDR tests ClanOv and ContextC.** Both tests rank domains (teal boxes) by *p*-value (numbers within boxes are ranks). **A.** In ClanOv ("Clan Overlap"), highest-ranking domains are considered as TPs, domains that overlap higher-ranking domains of the same clan (green connections) are removed (not counted toward the FDR or downstream overlaps), and domains that overlap higher-ranking domains of different clans (red connections) are considered FPs. **B.** In ContextC ("Context Coherence"), the highest-ranking domain in a sequence is considered a TP. Subsequent domains are considered TPs if there is at least one higher-ranking domain whose families have been observed together before (green connections), and otherwise they are considered FPs (all red connections).

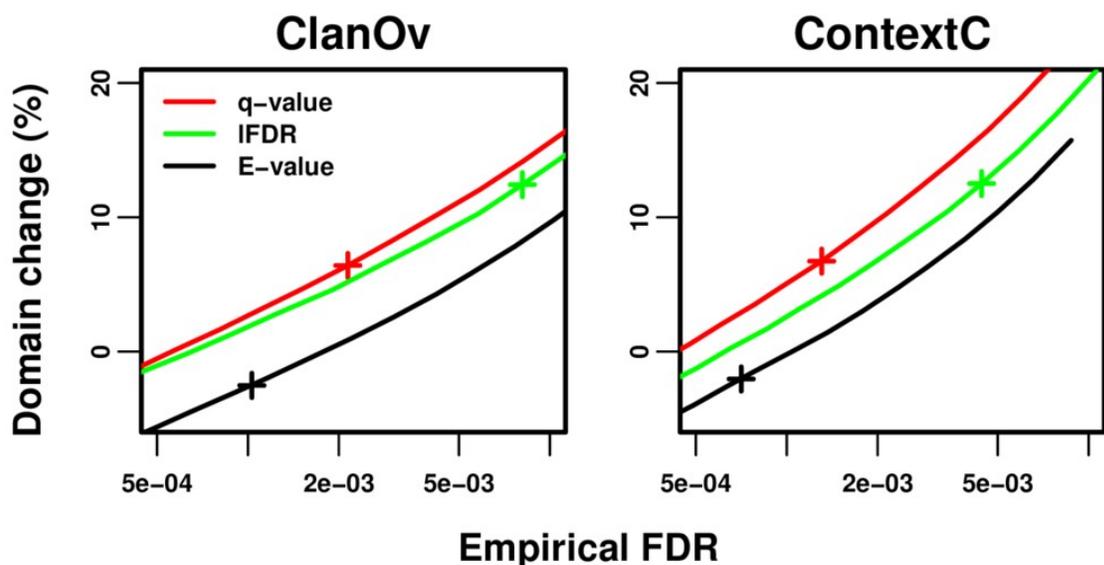

**Figure 3. Change in domain predictions while controlling empirical FDRs.** In each panel, a different empirical FDR test (*x*-axis, note log scale) is used to evaluate each method at a series of thresholds. The number of domain predictions is turned into a percent change relative to the number of Standard Pfam predictions (*y*-axis). Curves correspond to: *E*-values (black), cross marks $p \leq 1.3\text{e-}8$; stratified *q*-values (red), cross marks $q \leq 4\text{e-}4$; stratified lFDRs (green), cross marks $\text{lFDR} \leq 2.5\text{e-}2$. The *q*-value and lFDR thresholds marked with crosses correspond to the median of the Standard Pfam thresholds mapped theoretically to those statistics (see **Supplementary Results**). All curves have standard error bars in both dimensions, which are not always visible. Standard Pfam is not plotted as both the ClanOv and ContextC tests are based on Standard Pfam predictions.

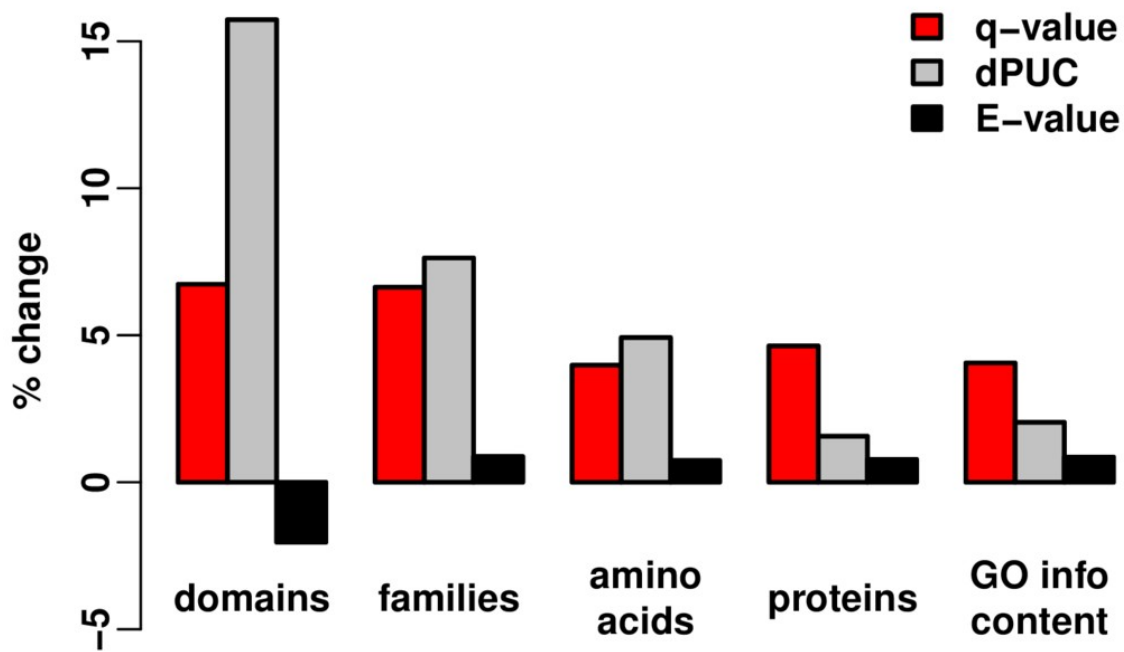

**Figure 4. Percent changes for several metrics relative to the Standard Pfam.** "Domains" counts domain predictions in UniRef50; "families" counts unique families per protein over all proteins; "amino acids" counts amino acids covered by domains over all proteins, without double-counting amino acids covered by multiple domains; "proteins" counts proteins with any predicted domain; "GO info content" sums the functional information content of all proteins with domain predictions (see **Supplementary Methods**). All quantities are turned into percent changes relative to the respective numbers from the Standard Pfam. *Q*-value uses *q*<=4e-4, *E*-value uses *p*<=1.3e-8, and dPUC uses a "candidate domain *p*-value threshold" of 1e-4, which gives comparable empirical FDRs.

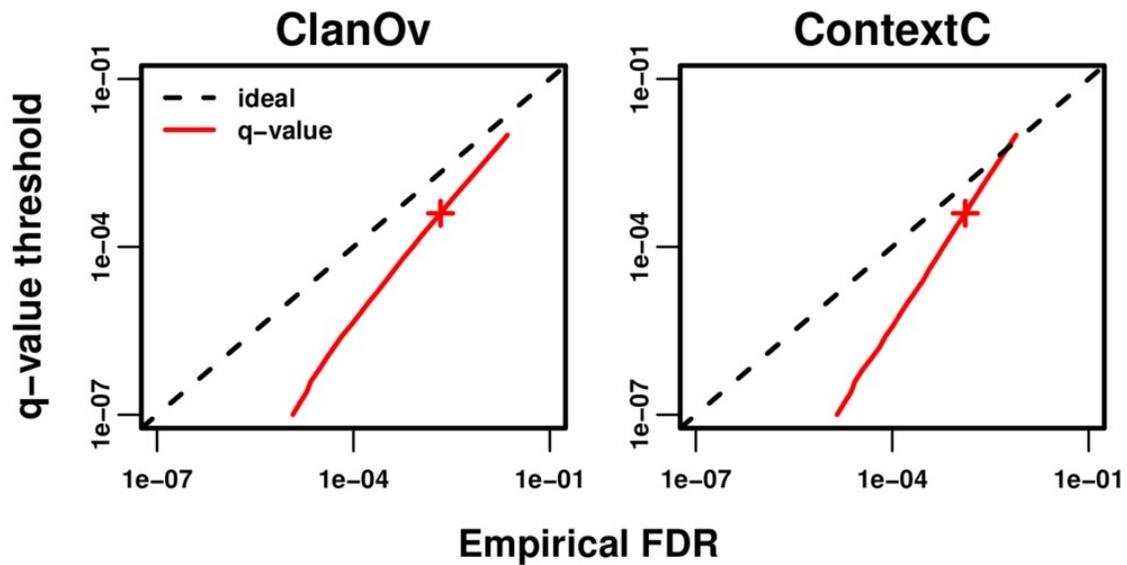

**Figure 5. Comparison of *q*-value thresholds and empirical FDRs.** In each panel, for each *q*-value threshold, observed empirical FDRs are computed (by ClanOv, left, and ContextC, right) and the relationship between these two quantities is shown in red. Since *q*-values control FDRs when input *p*-values are correct, ideally these data fall on the y=x line (dashed black lines). Values below the dashed line correspond to empirical FDRs that are larger than *q*-values. Smaller FDRs correspond to more stringent predictions and therefore include fewer predictions. All *x* and *y*-axes have the same range for ease of comparison and are in log scale. The red cross marks *q*<=4e-4.

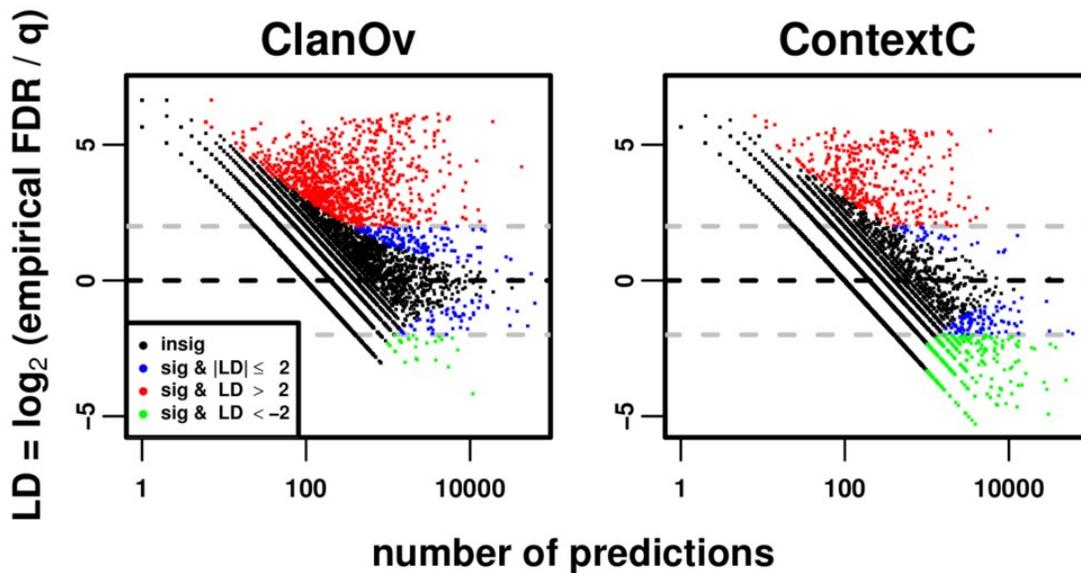

**Figure 6. Identification of domain families with empirical FDRs that differ significantly from expectation.** In each panel, the empirical FDR is computed (by ClanOv, left, and ContextC, right) for each domain family at q<1e-2, and the log-deviation (LD) of this empirical FDR from the threshold $q$=1e-2 is plotted on the y-axis relative to the number of predictions at this threshold (x-axis). Zero LD corresponds to perfect agreement (no deviation), while positive and negative numbers correspond to underestimated and overestimated empirical FDRs, respectively. The LD values of 0, 2, and -2 are marked with horizontal black, gray, and gray dashed lines respectively. Families are plotted as black dots if their deviations are insignificant via a Poisson test (**Methods**), blue if the deviations are significant but the effect size is small (|LD| <= 2), red if the deviations are significant and have a large positive effect size (LD > 2), and green if the deviations are significant and have a large negative effect size (LD < -2).

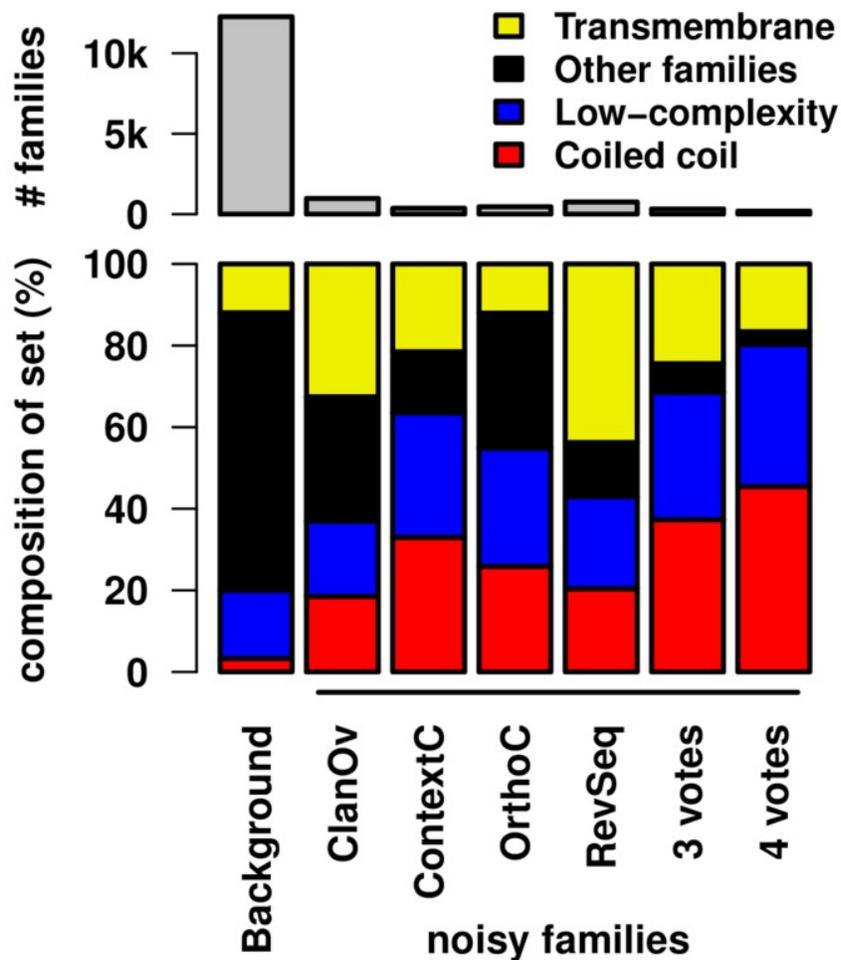

**Figure 7. Families with increased noise are enriched for repetitive patterns.** Each Pfam domain family was classified into one of four categories: transmembrane domain, low-complexity region, coiled-coil, and "other" (**Methods**). The bars for "background" corresponds to the set of all families in Pfam, and the rest of the bars correspond to the sets of noisy families identified as those with significantly larger empirical FDRs thatn expected with $q<=1e-2$ using either ClanOv, ContextC, OrthoC, RevSeq, at least three of these sets (3 votes), or all four tests (4 votes). The top (gray) bars show the size of each of these sets, and the bottom bars (colors) show the composition of these sets with respect to the four categories of domains. For each set of families, the significance of overlap with respect to each category was computed using the hypergeometric distribution. Two-sided $p$-values with $p<0.01$ were declared significant. All noisy sets were significantly enriched for coiled-coils and de-enriched for "other" families. Low-complexity regions were significantly enriched in all sets except ClanOv. Transmembrane domains were significantly enriched in all sets except OrthoC and "4 votes."

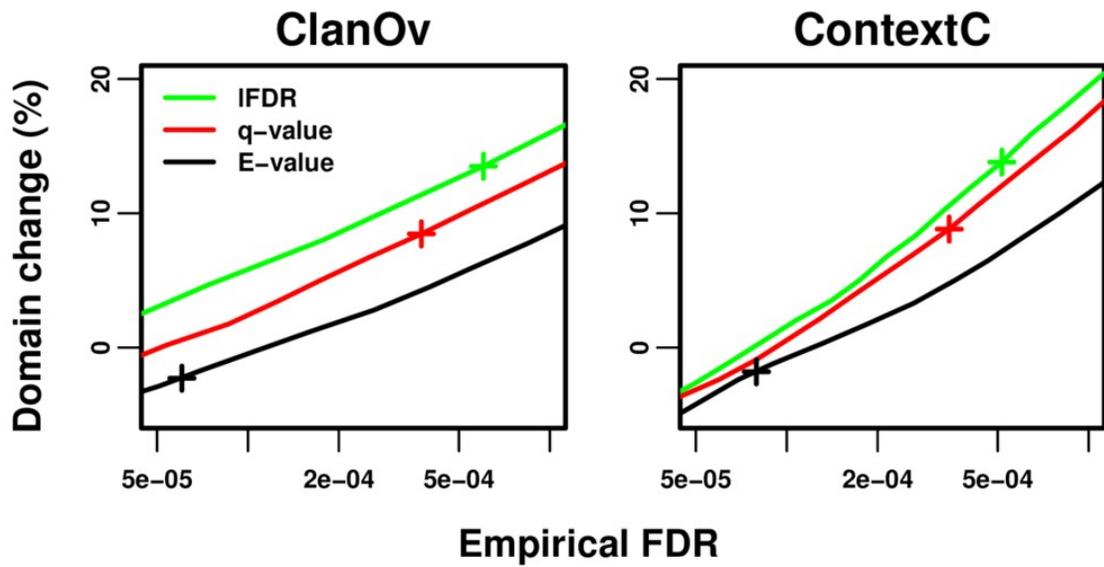

**Figure 8. Domain change while controlling empirical FDRs, restricted to families with as-expected noise.** This figure is exactly like **Figure 3** except that only families with "as-expected noise", rather than all families, are used in the benchmarks. See **Figure 3** for more information.